\documentclass[a4paper,fleqn,usenatbib]{mnras}

\usepackage{newtxtext,newtxmath}

\usepackage[T1]{fontenc}
\usepackage{ae,aecompl}

\usepackage{graphicx}	
\usepackage{amsmath}	
\usepackage{amssymb}	

\usepackage[flushleft]{threeparttable}

\newcommand{\so}{4U2206$+$54 }

\newcommand{\rxte}{{\it RXTE }}

\newcommand{\nustar}{{\it NuSTAR }}
\newcommand{\xmm}{{\it XMM-Newton }}
\newcommand{\sax}{{\it Beppo SAX }}
\newcommand{\integral}{{\it INTEGRAL }}

\title[No cyclotron line in 4U2206$+$54]{\nustar rules out a 
  cyclotron line in the accreting magnetar candidate 4U2206$+$54.}

\author[J.M. Torrej\'on et al.]{
J. M. Torrej\'on$^{1}$\thanks{E-mail: jmt@ua.es},
P. Reig$^{2,3}$,
F. F\"urst$^{4}$,
M. Martinez-Chicharro$^{1}$,
K. Postnov$^{5,6}$ 
\newauthor and L. Oskinova$^{7}$
\\
$^{1}$Instituto Universitario de F\'isica Aplicada a las Ciencias y las
Tecnlog\'ias, Universidad de Alicante, E-03690 Alicante, Spain\\
$^{2}$IESL, Foundation for Research and Technology-Hellas, 71110
Heraklion, Greece\\
$^{3}$ Physics Department and Institute of Theoretical and
Computational Physics, University of Crete, 70013, Heraklion, Greece\\
$^{4}$European Space Astronomy Center (ESAC), Science Operations
Department, E-28629 Villanueva de la Ca\~nada, Madrid, Spain\\
$^{5}$Sternberg Astronomical Institute, Moscow M.V. Lomonosov State
University, 119234 Moscow, Russia\\
$^{6}$National Research University Higher School of Economics, Myasnitskaya ul. 20, Moscow 101000, Russia\\
$^{7}$Institute for Physiscs and Astronomy, Universit\"at Potsdam,
D-14476 Potsdam, Germany
}

\date{Accepted XXX. Received YYY; in original form ZZZ}

\pubyear{2018}

\begin{document}
\label{firstpage}
\pagerange{\pageref{firstpage}--\pageref{lastpage}}
\maketitle

\begin{abstract}
Based on our new {\em NuSTAR} X-ray telescope data, 
we rule out any cyclotron line up to 60 keV in the
spectra of the high mass X-ray binary 4U2206$+$54. In particular, we do not find any evidence of
the previously claimed line around 30 keV, independently
of the source flux, along the spin pulse. The spin period has increased significantly, since
the last observation, up to $5750\pm10$ s, confirming the rapid
spin down rate $\dot \nu=-1.8\times 10^{-14}$~Hz s$^{-1}$. This behaviour
might be explained by the presence of a strongly magnetized neutron star
($B_{\rm s}>$ several times $10^{13}$ G) accreting from the
slow wind of its main sequence O9.5 companion. 
\end{abstract}




\section{Introduction}

The existence of accreting magnetars is an open 
question of modern Astrophysics. Magnetars are powered by magnetic energy. These
isolated neutron stars harbour the strongest cosmic magnets with field
strengths in the  $10^{13-15}$ G range \citep{1993ApJ...408..194T}. A wide range of high-energy 
phenomena displayed by Soft Gamma-ray Repeaters and Anomalous X-Ray 
Pulsars are explained by the extreme physics of magnetars
\citep{2002ApJ...574..332T, 2005ApJ...634..565T,
  2006csxs.book..547W}. About 30 magnetars and candidates are now
known \citep{2014ApJS..212....6O}, but none of them is an
\emph{accreting} neutron star (NS). 

An NS is a remnant of a massive star with $M_{\rm initial}>8\,M_{\odot}$. 
The vast majority of massive stars are binaries
\citep{2012MNRAS.424.1925C, 2012Sci...337..444S}. Only a small
fraction of these binaries remain bound after 
the primary explodes as a supernova but the number of such systems in the 
Galaxy is still quite large \citep{2006A&A...455.1165L}. The accretion of the stellar 
wind onto the NS powers a strong X-ray luminosity 
in these high-mass X-ray binaries (HMXBs). In our current 
understanding on how binary systems evolve
\citep[e.g.][]{1973A&A....25..387V, 2014LRR....17....3P}, and given the
  magnetic field strength distribution of known pulsars \citep[][Fig.7]{2014ApJS..212....6O}, a 
(small) fraction of HMXBs should host neutron stars with magnetar strength
fields, i.e., being accreting magnetars. Yet, the very existence of the class
is unclear. 

A handful of accreting magnetar candidates have now been proposed,
among them the long period X-ray pulsars 4U0114$+$65
\citep{2017A&A...606A.145S} and AX J1910.7$+$0917 \citep{2017MNRAS.469.3056S}. The magnetar nature of our target,  4U2206$+$54, has been suggested on the grounds of its very 
long spin period ($P_{s}=5560$ s, the third largest known after AX
J1910.7$+$0917 and 4U0114$+$65) and a very high
period derivative, $\dot{\nu}=-1.5\pm0.2 \times 10^{-14}$ Hz s$^{-1}$
\citep{2010ApJ...709.1249F, 2012MNRAS.425..595R} which would drive
  the NS to a complete halt in $\Delta t\gtrsim 300$ yr.  The possible
  explanations involve NS spin magneto-braking in a magnetic
field exceeding  $B\sim 10^{13}$ G \citep[e.g.][]{2010Ap.....53..237I}.



However, the magnetar nature of the neutron star in 4U2206$+$54 has been 
disputed because the extant low signal to noise X-ray spectra 
do not rule out the presence of a cyclotron resonant scattering
feature (CRSF; from now on, a cyclotron line) at $E_{\rm cyc}\approx
29$ keV, a clear signature of a magnetic 
field of $B=3.3\times 10^{12}$ G, definitely not in the magnetar range. Although
at a quite low significance, the 
detection has been claimed on the basis of spectra obtained with several 
satellites (\sax, \rxte, \integral) and at different epochs
\citep[][]{2004A&A...423..301T, 2005A&A...438..963B,
  2009MNRAS.398.1428W}. If this line is indeed present in the X-ray spectra,
the system contains no magnetar. Further constraints have then
to be added in order to accomodate the pronounced spin
down. \cite{2013ARep...57..287I} propose the magnetic
accretion model. This mechanism requires a (low) magnetization of the donor
stellar wind. In such a case, a dense magnetic slab of ambient matter forms around the
magnetosphere of the NS, able to remove the angular momentum at the
observed rate even for normal strength magnetic fields ($B \sim
10^{12}$ G). However, a joint analysis of (non
contemporaneous) \xmm and \integral data, did not reveal the presence
of any cyclotron line \citep{2012MNRAS.425..595R}, casting serious
doubts on its existence. 

The goal of this paper is to confirm or rule out such a line. To
that end we present an analysis of a 58.6 ks \nustar observation of \so
covering, for the first time, the energy range from 3 to 60 keV with
no spectral gaps. The high S/N ratio provides a stringent test to the presence of a cyclotron line. 

The paper is structured as follows: in Section \ref{sec:obs} we
present the observational details. In sections \ref{sec:lc} and
\ref{sec:spec} we analyze the light curve and flux resolved spectra of
the source, providing the best fit parameters for the continuum. Finally, in Sections \ref{sec:disc} and
\ref{sec:conc} we discuss these parameters in the framework of the
theory and present the conclusions.

\section{Observations}
\label{sec:obs}

\begin{table}
	\centering
	\caption{Observations journal}
\begin{threeparttable}
	\label{tab:obs}
	\begin{tabular}{lccr} 
		\hline
		ObsID & Date & $t_{\rm exp}$ & $\phi _{\rm orb}$ \\
		\hline
		30201015002 &  2016-05-17 09:51:08\tnote{a} & 58.6 &
                0.03 (0.22)\tnote{b}\\
		\hline
	\end{tabular}
\begin{tablenotes}
\item[a] MJD 57525.41050926
\item[b] Time for X-ray maximum $T_{0}= 51856.6\pm 0.1$ MJD
  \citep{2006A&A...449..687R}. Phases for orbital periods
$P=9.5591\pm 0.0007$ d  and 
$P=19.25\pm 0.08$ d \cite{2007ApJ...655..458C}, respectively.

\end{tablenotes}
\end{threeparttable}
\end{table}

We observed \so with \nustar on 17 May 2016 during 58.6 ks on target. In Table
\ref{tab:obs} we specify the observation journal details. The \nustar
data were extracted with the standard software \texttt{nupipeline}
v1.7.1, which is part of HEASOFT v6.20. We used CALDB version
20170222. The data were extracted from a circular region with radius
of 100$''$ centred on the brightest pixels in the image after standard
screening. The background was extracted from a circular region with a
radius of 150$''$ located as far away from the source as possible
within the field of view. We also extracted data from the
\texttt{SCIENCE\_SC} mode, during which the pointing is less precise
\citep[see, e.g.][]{2016ApJ...826...87W}, adding about 15\% exposure time. We
used a circular region with $90''$ radius to extract the source
spectra and lightcurves from these data and a similar background
region as in the standard data. All data were corrected for the solar system barycentered using the DE200 ephemeris. The
spectral analysis was performed with the Interactive Spectral Interpretation
System (\textsc{isis}) v 1.6.1-24 \citep{2000ASPC..216..591H}.

\section{Light curves and timing analysis}
\label{sec:lc} 

\begin{figure}
\resizebox{\hsize}{!}{\includegraphics{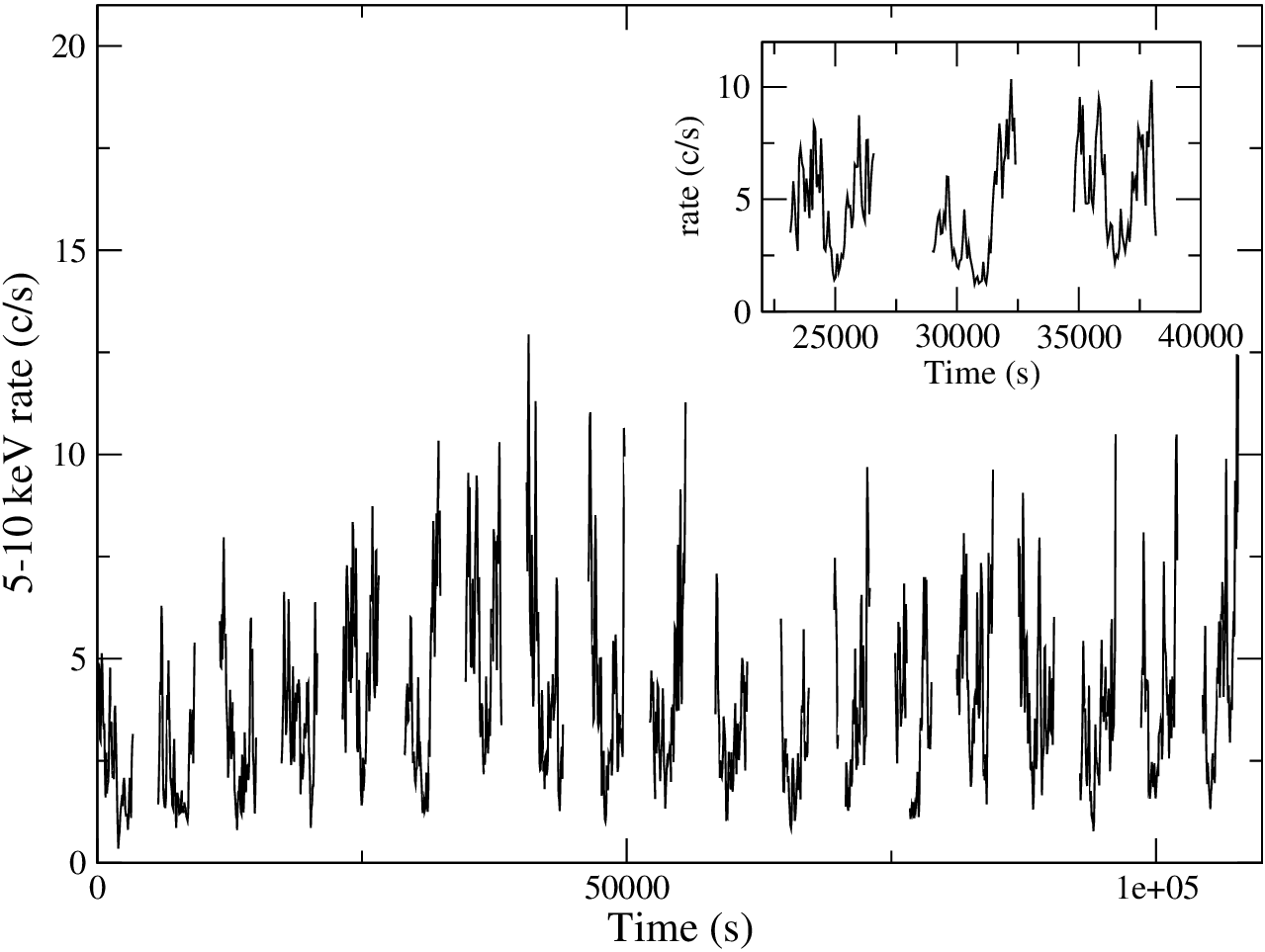}}
\caption[]{$5-10$ keV light curve of 4U\,2204+54. The inset shows a more detailed
view of the variability. Strong evidence for pulsations is seen. Time zero
corresponds to JD 2,457,525.912}
\label{fig:lc}
\end{figure}

We extracted energy resolved lightcurves between 3--5\,keV, 5--10\,keV, and
10--79\,keV with 10\,s time resolution and a broadband lightcurve between
3--79\,keV with 1\,s time resolution.  Figure~\ref{fig:lc} shows the \nustar
background subtracted light curve in the energy range 5-10 keV with a bin size
of 60 s. In this energy band, the source shows the
highest count rate while the effects of  photoelectric absorption are
minimized. The light curve is strongly variable as typical in wind
accreting HMXBs. This stochastic variation is superimposed to the pulse of the
NS. The relatively large amplitude variations are due to the pulsations, as
can be seen in the inset.

\begin{table}
\begin{center}
\caption{Spin period determination. }
\label{period}
\begin{tabular}{lc}
\hline \hline \noalign{\smallskip}  
Method		&$P_{\rm spin}$ (s)	\\
\hline \noalign{\smallskip}
\multicolumn{2}{c}{Fundamental}\\
\hline
Lomb-Scargle	&$5750\pm10$   \\
PDM		&$5741\pm38$  \\
CLEAN		&$5752\pm38$	\\
CHISQ		&$5740\pm68$	\\
\hline
Mean			&5746$\pm$6	\\
Weighted mean		&5752$\pm$9	\\
\hline
Final adopted value	&5750$\pm$10	\\
\hline \hline \noalign{\smallskip}
\end{tabular}
\end{center}
\end{table}

The determination of the spin period in 4U\,2204+54 with {\it NuSTAR} data is
difficult owing to the structure of the light curve and the sampling of the
periodicity.  Due to the low-Earth orbit, {\it NuSTAR} data are affected by the
passage near the South Atlantic Anomaly and suffers from Earth
occultations.
Therefore the light curve contains numerous gaps interspersed with continue data
segments. The observation spans over $\sim$108 ks, but the total on-source time
was 58.6 ks. 
Although the data segments are evenly sampled (bin
size of 10 s),  the spin period of 4U\,2204+54 is longer than the typical
duration of each data segment. In other words, the phase coverage of the
observed signal is incomplete.

Under these conditions, the use of Fourier techniques is inappropriate. The good
news is that  we know the value of the periodicity, hence we do not need to
perform a blind search and we can restrict the relevant frequency range.  The
latest reported value of the spin period of 4U\,2204+54 is 5593 s, obtained from
an \xmm observation in 2011  \citep{2012MNRAS.425..595R}. They  
derived a spin-down rate of $(1.5\pm0.2) \times 10^{-14}$ Hz s$^{-1}$. If we
assume that this rate continued until the {\it NuSTAR} observation on JD
2457525.91, then the expected period would be $\sim 5690$ s. We note that
the orbital period of the satellite is 5828 s. 

We based our analysis on the the Lomb-Scargle periodogram (LSP)
\citep{1976Ap&SS..39..447L,1982ApJ...263..835S}.  There are various
implementations of the LSP. The differences appear in the normalization of the
periodogram, whether  the zero point of the sinusoid is allowed to change during
the fit \citep{1999ApJ...526..890C}, the treatment of errors
\citep{2009A&A...496..577Z}, and in the computation speed
\citep{1989ApJ...338..277P, 2010ApJS..191..247T}.

\begin{figure}
\includegraphics[width=0.75\columnwidth, angle=0]{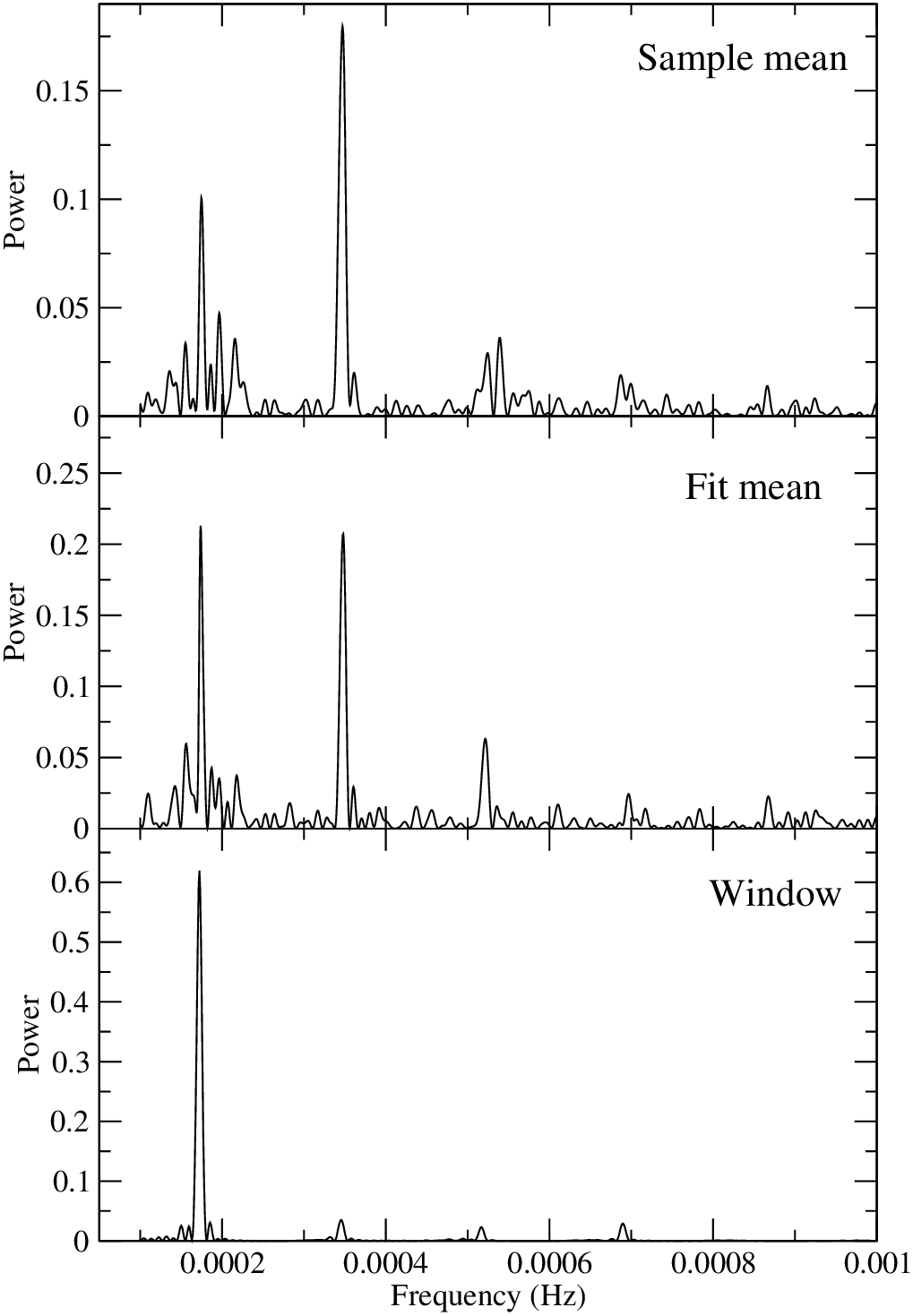}
\caption[]{Lomb-Scargle periodograms of the 5--10 keV light curve of 4U\,2204+54
using different implementations of the algorithm. {\it Top:} sample mean as in
the original Lomb-Scargle periodogram. {\it Middle:} Python implementation
allowing the model fits for the mean of the data. {\it Bottom} Power spectrum of 
the window function.}
\label{LS}
\end{figure}

The upper panel of Fig.~\ref{LS} shows the LSP using the original formulation of
pre-centering the data to the sample mean. Two peaks are apparent. The first
peak at $\nu=0.0001739$ Hz and the second peak at twice that frequency. 
Although the first frequency is close to the expected period, its power is
significantly lower than the second peak.  The reason for the suppression of the
power of the fundamental peak is the use of the sample mean in a time series
when the data do not provide full phase coverage of the observed signal
\citep{2017arXiv170309824V}. In the middle panel of Fig.~\ref{LS} we show the
LSP using the floating mean method, which involves adding an offset term to the
sinusoidal model at each frequency \citep{1999ApJ...526..890C,
2009A&A...496..577Z}. The average spin period obtained from the LSP is
$5750\pm10$ s. The error was estimated from the dispersion of the different
values obtained after running the various implementations of the Lomb-Scargle
periodogram.
We confirmed the high significance of the peak by calculating the
false alarm probability (the detection threshold above which the signal is
significant) through bootstrap analysis. We find that the peak is significant
well above 99.99\%.


\begin{figure}
\resizebox{\hsize}{!}{\includegraphics{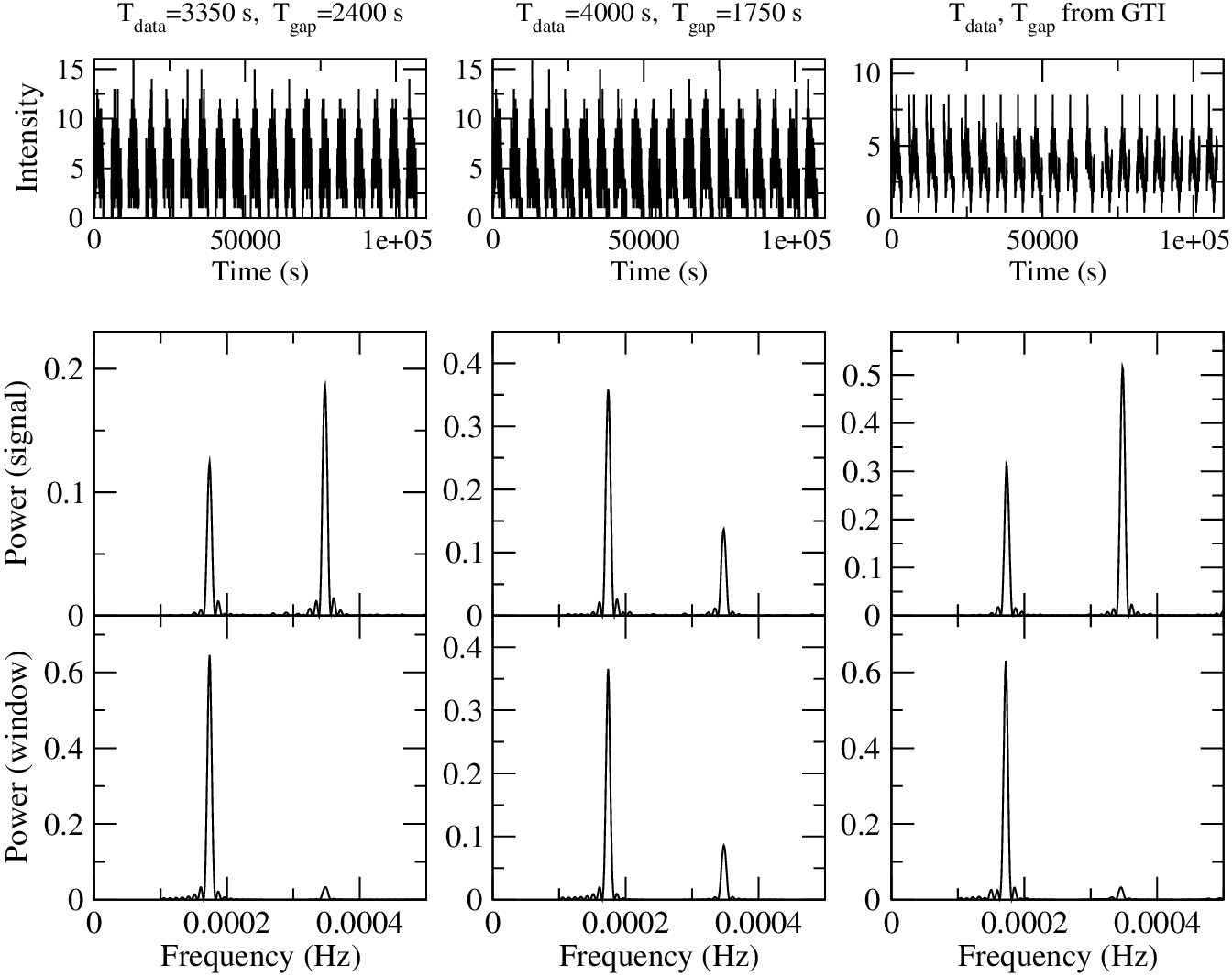}}
\caption[]{Simulated light curve with $P=5750$ s (upper panels). Below
  each lightcurve, the LSP 
and window spectrum for different selections of $T_{\rm data}$ 
and $T_{\rm gaps}$, are shown. See text for details.}
\label{simul}
\end{figure}

To examine the conditions under which the suppression of the main peak
occurs, we simulated a purely sinusoidal signal affected by Poisson noise with a
period of $P_{\rm spin}=5750$ s. Then we removed data points at regular
intervals, each interval with a duration of $T_{\rm gaps}$, leaving continuous
stretches of data of duration $T_{\rm data}$. This is illustrated in
Fig.~\ref{simul}, where two cases which differ in the value of  $T_{\rm data}$
and $T_{\rm gaps}$ are shown. In both cases we assume that $T_{\rm data}+T_{\rm
gaps}=P_{\rm spin}$.

In the left and middle columns of Fig.~\ref{simul}, $T_{\rm data}=3350$ s,
$T_{\rm gaps}= 2400$ s and $T_{\rm data}=4000$ s, $T_{\rm
gaps}= 1750$ s, respectively. A strong peak is apparent in the window spectrum at a frequency 
$1/(T_{\rm data}+T_{\rm gaps})$, which in this case is equal to $1/P_{\rm
spin}$. When the data segments do not cover a good fraction of the pulse phase,
$T_{\rm data}/P_{\rm spin}=0.6$, the power of the peak that corresponds to the
true period is suppressed (left column, signal panel). As the phase coverage increases
($T_{\rm data}/P_{\rm spin}$ approaches 1), the power of the true period
increases (middle column, signal panel). 

We also performed a simulation using the exact exposure window of the
observations (right column in Fig.~\ref{simul}). In this case, the light curve
was created by replicating the pulse profile  to cover the length of the
observation. Poisson noise was added to each bin. The pulse profile was obtained
using the derived period of 5750 s. Then we introduced gaps using the same good
time intervals as the real light curve.  The first peak appears at the expected
frequency.  In the real light curve and the simulated one using the real good
time intervals, the peak of the window power spectrum appears at $\nu=0.0001721$
Hz ($P=5811$ s), while the spin period occurs at  $\nu=0.0001739$ Hz ($P=5750$
s). These simulations give us confidence that the peak at frequency $\sim$0.0001739
Hz in the LSP of 4U\,2204+54 is real and corresponds to the true
period.

%
\begin{figure}
\resizebox{\hsize}{!}{\includegraphics{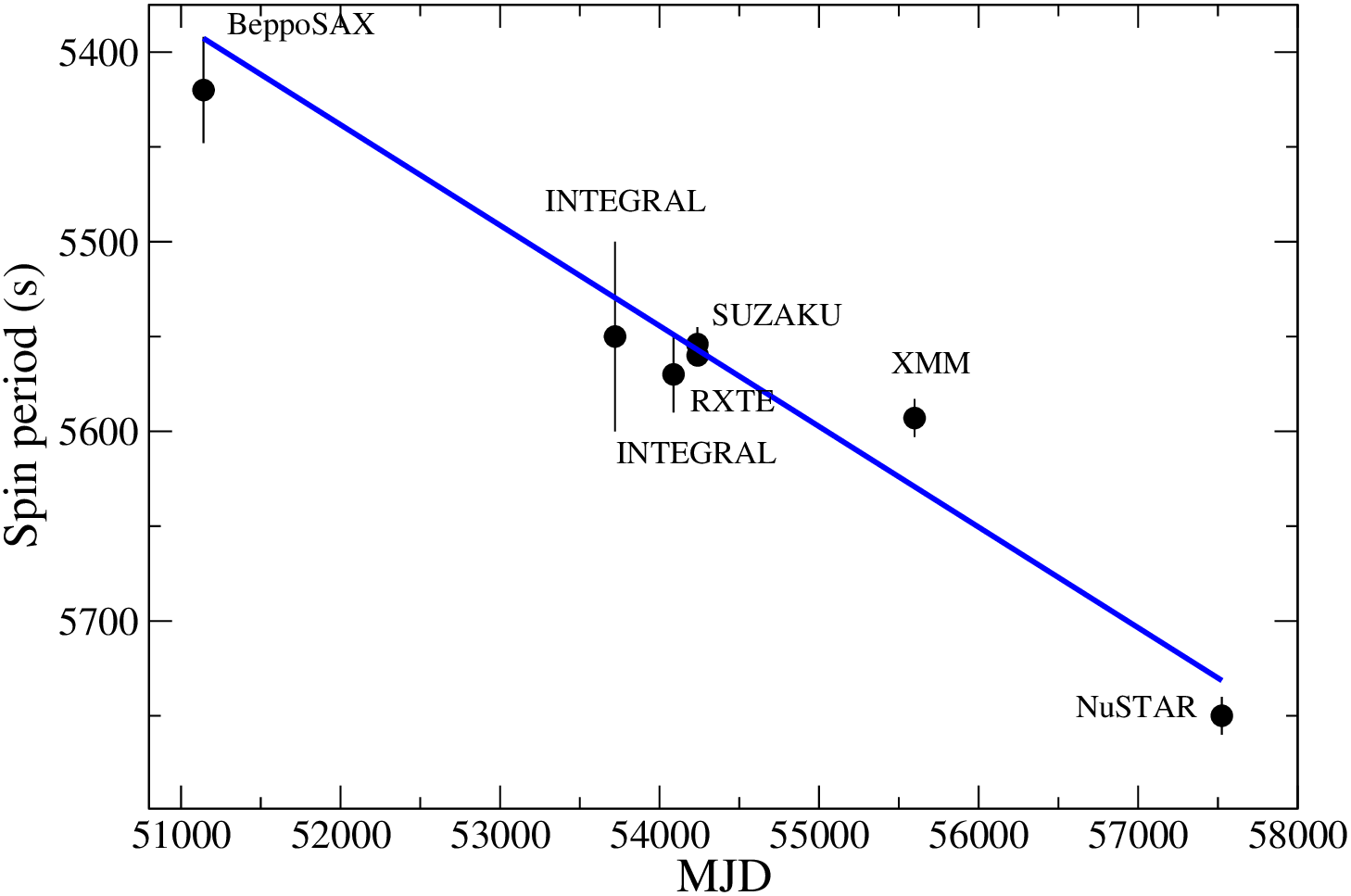}}
\caption[]{Spin period evolution of 4U\,2204+54. The best linear fit is $P_{\rm
spin}=2680+0.053 * MJD$, with errors $\sigma_y=160$ and $\sigma_x=0.003$. }
\label{spinevol}
\end{figure}

We compared the average spin period obtained from the LSP  value with those
obtained using other techniques such as the \textsc{clean}
\citep{1987AJ.....93..968R}, the phase dispersion minimization
\citep[\textsc{pdm}][]{1978ApJ...224..953S} algorithms and $\chi^2$
minimization. These algorithms are implemented in the program PERIOD (version
5.0), distributed with the \textsc{starlink} Software Collection. Note that the
final "clean" spectrum of the \textsc{clean} deconvolution algorithm detects
only the first harmonic. The reason is that the \textsc{clean} algorithm assumes
that the highest peak in the periodogram corresponds to the primary signal.  
Table~\ref{period} summarizes the results of the period search.  We took the
result from the LSP as the final adopted value of the spin period of
4U\,2204+54. The timing analysis was performed using the 5--10 keV light curve.
However, the analysis performed at other energy ranges gave consistent results.

Figure~\ref{spinevol} shows the evolution of the spin period of 4U\,2204+54 over
the past 20 years. Data prior to the {\it NuSTAR} observation were taken from
\citet{2012MNRAS.425..595R}. A linear fit to the data represents a good approximation of the
long-term variation of the spin period with time. The source continues to spin
down at a rate of $(1.8\pm0.1) \times 10^{-14}$ Hz s$^{-1}$. This value agrees
with previous reported values $(1.7\pm0.3) \times 10^{-14}$ Hz s$^{-1}$
\citep{2010ApJ...709.1249F} and  $(1.5\pm0.2) \times 10^{-14}$ Hz
s$^{-1}$ \citep{2012MNRAS.425..595R}.

\begin{figure}
\includegraphics[angle=0,width=1\columnwidth]{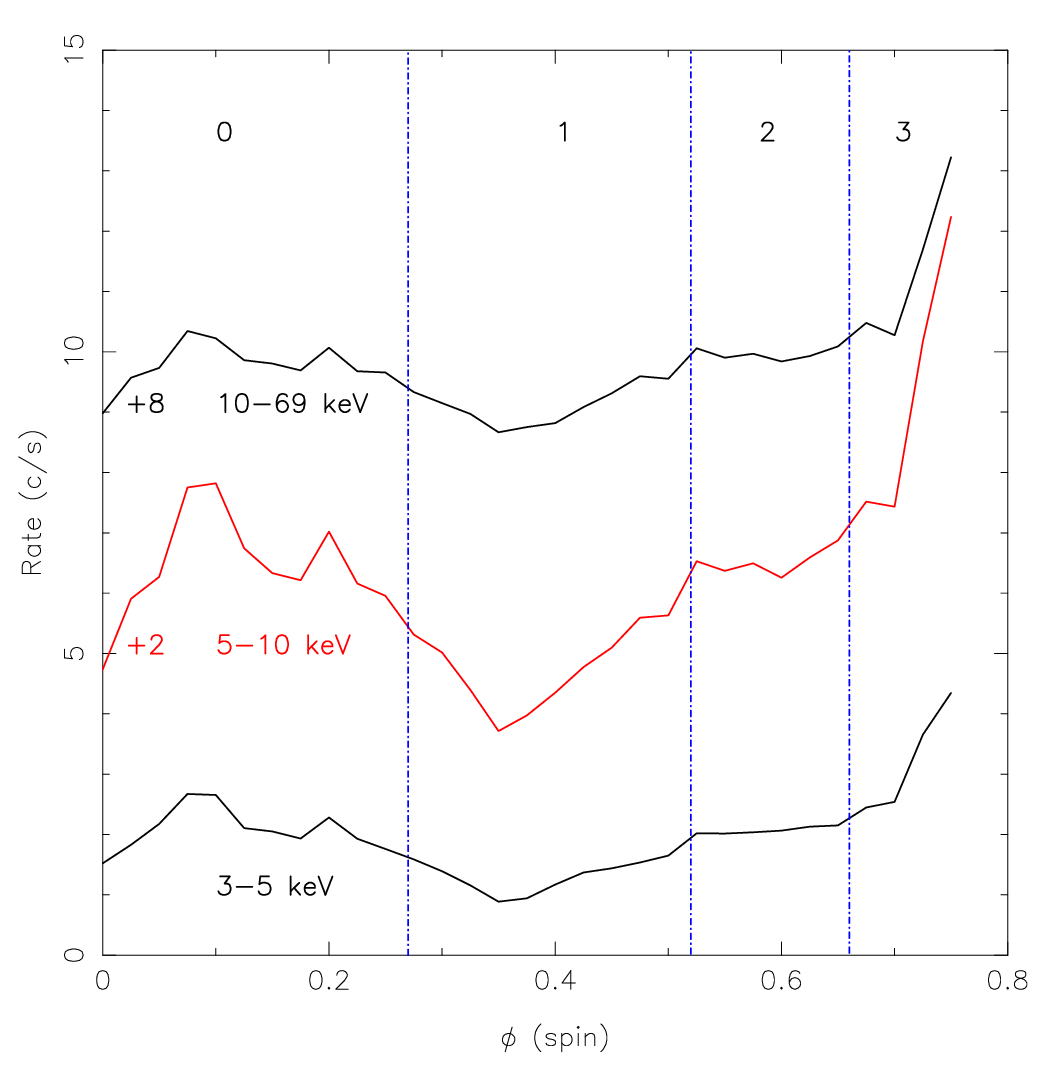}
\caption{Average spin pulse, as a function of energy, folded over the  $P_{\rm
  spin}=5750\pm10$ s period. Offsets of $+2$ and $+8$ c s$^{-1}$ has
been added, for the sake of clarity. Several flux levels of the source are
separated by vertical lines. Note the similarity with the individual pulse in
the inset of Fig.~\ref{fig:lc}.}
\label {fig:phascuts}
\end{figure}

In Fig, \ref{fig:phascuts} we show the average spin pulse folded over the
\nustar period, $5750\pm10$ s, for several energy ranges. It shows a double peak structure. Several
emission flux levels are identified and separated by vertical lines. These will
be used to perform flux resolved spectroscopy in the next section.

\section{Spectra}
\label{sec:spec}

\begin{figure}
\includegraphics[angle=0,width=1.0\columnwidth]{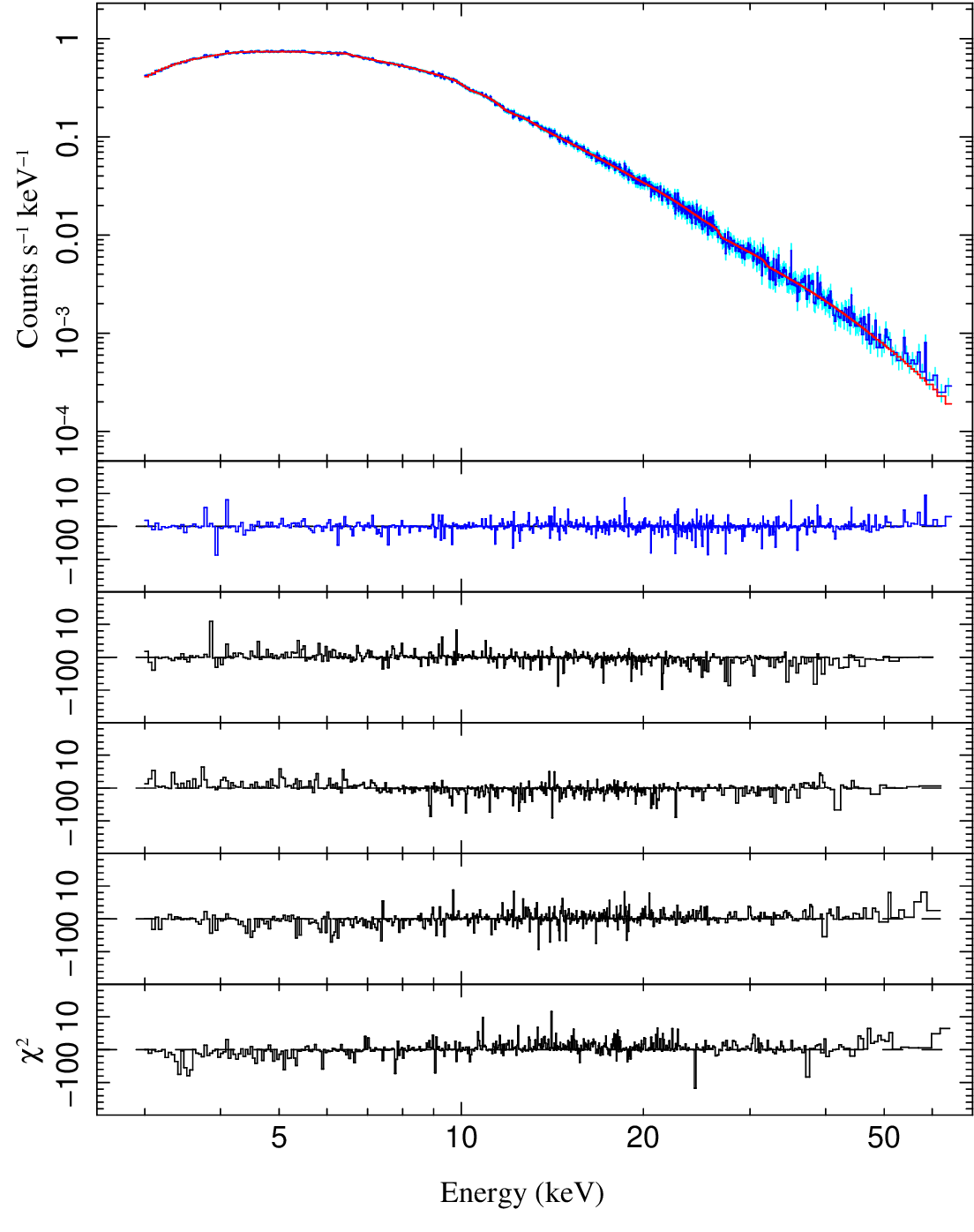}
\caption{\nustar average spectrum (blue line) and best fit model (red line). The spectral models are
described in detail in Section \ref{sec:spec} and parameters given in
Table \ref{tab:continuum_compmag}. The residuals
correspond to average (top) and spin phases 0, 1, 2 and 3, respectively.}
\label {fig:f+q_spec}
\end{figure}

The observed X-ray spectra (blue) and the best fit model (red) are
presented in Fig. \ref{fig:f+q_spec}. In order to search for cyclotron lines at high energies, a good fit of
the underlying continuum is required. The X-ray continuum of \so has
been satisfactorily described, in the past, using
bulk motion comptonization models \citep[for \rxte, \sax and \xmm
respectively]{2004A&A...423..301T, 2012MNRAS.425..595R}. Our \nustar
spectra cover, for the first time, the energy interval 3-60 keV
uninterrupted. The lack of spectral gaps, which eliminates any
uncertainty in inter-calibration constants, together with
the high S/N ratio at energies $E>20$ keV, allows us
to constrain the model parameters with high accuracy. The continuum is well described by the hybrid thermal and dynamical (bulk)
componization model \textsc{compmag} \citep{2012A&A...538A..67F}. 

\textsc{compmag} is a 
model for comptonization of soft photons which solves the
radiative transfer equation for the case of cylindrical accretion onto
a magnetised NS. 
The soft seed
photons, with temperature $kT_{\rm bb}$ are upscattered by the
infalling plasma
in the accretion column, with temperature $kT_{\rm e}$. In order
  to gain stability for the computation of the uncertainties, some of
  the model parameters must be kept fixed during the fits. After a number
of tests, the best results were achieved by setting the infalling plasma
velocity increasing towards the NS surface ($\beta$ flag 1) with a
velocity law index $\eta$ compatible with 1 in all cases
\citep[see][for full details and
references]{2012A&A...538A..67F}. The spectra show little
  variations across the spin pulse and the vast majority of
  parameters (except the normalizaion) are compatible within the errors.

The emitted X-ray continuum described above fits the high energy spectrum
perfectly. Below 6 keV, however, the model falls below the data even
for $N_{\rm H}=0$. This \emph{soft excess} is a well known feature of
many HMXBs \citep{2004ApJ...614..881H} and its origin is still
unclear. To fit it, we added a blackbody with a
temperature equal to that of the soft photons source $kT_{\rm
  bb}$. Although the general $\chi^{2}_{r}$ was acceptable, some
residuals clearly remain at low energies. The best fit is achieved
when both temperatures are decoupled. The temperature of the
additional blackbody turns out to be half that of the soft seed
photons. 

Both components are modified at low energies by photoelectric absorption which
accounts for the local and interstellar material. The photoelectric absorption has been modeled using
\texttt{tbnew} which contains the most up to date cross sections for
X-ray
absorption\footnote{http://pulsar.sternwarte.uni-erlangen.de/wilms/research/tbabs\\/index.html}. Finally,
a gaussian component has been added to describe the Fe K$\alpha$
fluorescence line. The
best fit parameters are presented in Table
\ref{tab:continuum_compmag}. The 
norm of the additional blackbody is 5 orders of magnitude lower than the {\texttt compmag}
component. In any case, it has no impact on the high energies we are
focused in, and we do not discuss it further.

In agreement with previous studies, the temperature of the soft photon
source is quite high $kT_{\rm bb}\simeq 1.5$ keV. At a distance
  of $d=3.7\pm0.4$ kpc \citep[][]{2016A&A...595A...1G, 2018arXiv180409376L} the
0.3--20 keV luminosity would be in the range $L_{\rm X}=[4.7-11]\times
10^{35}$ erg s$^{-1}$ for spin phases 1 (low flux) and 3 (high flux),
respectively ($6.5\times 10^{35}$ erg s$^{-1}$ average), roughly one order of magnitude lower than those usually
found in HMXBs. This is consistent with a small emission area,
presumably a hot spot on the NS surface
\citep{2004A&A...423..301T}. The conclusion is further supported by the \nustar analysis. Indeed, $r_{0}$ is the accretion column radius, in units of
the NS Schwarzschild radius, or $r_{0}\simeq 0.84$ km. On the other hand,
from the normalization constant
$(R_{\rm km}/d_{10\rm kpc})^{2}$ a soft photons source radius of  $R_{\rm km} =
1.7$ km is derived. Finally, the albedo, $0< A <1$, is consistent
with reflection from a NS surface\footnote{$A\approx 0$ is expected
  for a Black Hole} .


\begin{table*}
{\def\arraystretch{1.3}
\begin{center}
\caption{Model \textsc{compmag+bb} continuum parameters. Uncertainties
are given at 90\% confindence level. }
\begin{threeparttable}
\label{tab:continuum_compmag}
\begin{tabular}{ccccccc}
\hline\hline
Parameter &  $\phi_{\rm spin}=0$ & 1 & 2 & 3 & average\\
\hline
&&COMPMAG&&\\
$N_{\rm H,1}$ (10$^{22}$ cm$^{-2}$) & 0.8$^{+0.3}_{-0.3}$ & 0.9$^{+0.4}_{-0.4}$& 0.83$^{+0.15}_{-0.15}$& 0.83$^{+0.19}_{-0.19}$& 0.83$^{+0.09}_{-0.07}$\\

 Norm &23.96$^{+0.11}_{-0.31}$&15.58$^{+0.06}_{-0.06}$&25.03$^{+0.11}_{-0.09}$&34.88$^{+0.13}_{-0.13}$& 20.98$^{+0.04}_{-0.04}$\\

$kT_{\rm bb}$ (keV) &1.54$^{+0.04}_{-0.04}$&1.54$^{+0.01}_{-0.01}$&1.543$^{+0.002}_{-0.002}$&1.544$^{+0.002}_{-0.002}$& 1.557$^{+0.001}_{-0.001}$\\

$kT_{\rm e}$ (keV) & 16.0$^{+0.4}_{-0.5}$& 15.9$^{+0.6}_{-0.6}$& 17.00$^{+0.03}_{-0.24}$& 16.78$^{+0.21}_{-0.21}$& 16.01$^{+0.11}_{-0.10}$\\

$\tau$ & 0.23$^{+0.11}_{-0.11}$& 0.23$^{+0.10}_{-0.10}$& 0.225$^{+0.001}_{-0.001}$& 0.222$^{+0.002}_{-0.002}$& 0.226$^{+0.001}_{-0.001}$\\


$\beta_{0}$ &0.08$^{+0.09}_{-0.09}$&0.13$^{+0.07}_{-0.07}$&0.127$^{+0.002}_{-0.002}$&0.127$^{+0.017}_{-0.015}$&0.081$^{+0.011}_{-0.011}$\\

$r_{\rm 0}$ &0.2$^{+0.1}_{-0.1}$&0.2$^{+0.1}_{-0.2}$&0.195$^{+0.002}_{-0.001}$&0.198$^{+0.001}_{-0.001}$&0.200$^{+0.001}_{-0.001}$\\

$A$ &0.5$^{+0.3}_{-0.3}$&0.5$^{+0.3}_{-0.3}$&0.61$^{+0.02}_{-0.01}$&0.66$^{+0.02}_{-0.02}$&0.57$^{+0.01}_{-0.01}$\\


Flux\tnote{a} & $4.49^{+0.01}_{-0.01}$& $2.86^{+0.01}_{-0.01}$ & $4.73^{+0.02}_{-0.02}$ & $6.61^{+0.02}_{-0.02}$ & $3.94^{+0.09}_{-0.09}$ \\
&&&&\\

&&BB&&\\
 Norm (10$^{-4}$)&5.4$^{+0.3}_{-0.3}$&3.30$^{+0.12}_{-0.12}$&5.8$^{+0.3}_{-0.2}$&7.1$^{+0.4}_{-0.4}$&4.77$^{+0.11}_{-0.11}$\\

$kT_{\rm }$ (keV) &0.75$^{+0.27}_{-0.17}$&0.75$^{+0.13}_{-0.17}$&0.62$^{+0.01}_{-0.01}$&0.61$^{+0.01}_{-0.01}$&0.69$^{+0.01}_{-0.01}$\\

Flux\tnote{a} & 0.28$^{+0.02}_{-0.02}$&0.17 $^{+0.01}_{-0.01}$ &0.24 $^{+0.01}_{-0.01}$ & 0.28$^{+0.02}_{-0.02}$& 0.23$^{+0.01}_{-0.01}$ \\
&&&&\\
&&GAUSS&&\\

$E$  (keV)&6.42$^{+0.10}_{-0.10}$&6.42$^{+0.18}_{-0.06}$&6.42$^{+0.13}_{-0.13}$&6.42$^{+0.29}_{-0.29}$&6.42$^{+0.08}_{-0.06}$\\
Flux ($\times 10^{-6}$ ph s$^{-1}$ cm$^{-2}$) & 40$^{+42}_{-24}$& 43$^{+17}_{-20}$& 23$^{+26}_{-23}$& 29$^{+37}_{-30}$& 46$^{+23}_{-23}$\\


$EW$ (eV)& 4.3$^{+4.3}_{-2.4}$& 7.6$^{+3.1}_{-2.4}$ & 2.5$^{+2.8}_{-2.5}$& 2.5$^{+2.8}_{-2.3}$  & 5.91$^{+3.11}_{-3.11}$  \\

$\chi^{2}_{\rm r}$(d.o.f.) &  1.02(522)    & 0.82(510) & 1.14(532)& 1.04(529)& 0.98(745)\\
\hline\hline
\end{tabular}
\begin{tablenotes}

\item[a]{Unabsorbed $2-65$ keV flux $\times 10^{-10}$ erg s$^{-1}$
    cm$^{-2}$}
\item[]{$\beta$ flag and $\eta$ fixed to 1} 
\item[]{$\sigma_{\rm Fe}$  fixed to 0.12 keV}
\end{tablenotes}
\end{threeparttable}
\end{center}
}
\end{table*}

\subsection{No CRSF at 30 keV}

The contiuum described above fits the \nustar spectra of \so
perfectly in the high energy range. In
Fig. \ref{fig:f+q_spec} we show the phase averaged spectrum, the best
model and the corresponding residuals. The
region 20-40 keV does not show any
peculiar residuals and does not require the addition of the previously
claimed cyclotron line at $E_{\rm cyc}\sim 29$ keV. In order to
  search for any possible dependence on source brightness, we plot
  also the residuals of each individual spin phase, to the average
  model, renormalised. The spectrum shows little variability across
  the spin pulse. Likewise, no cyclotron line is detected in any of
  the spin phases. 

In Table \ref{tab:cyclobs} we compile the cyclotron line (C.L.) claims
along with the corresponding average source flux. The line is only
detected in some observations but, again, there seems to be no
 correlation with the source flux in the long term. This casts serious doubts on the
 existence of the line. Thus, we can
safely rule out the presence of any cyclotron line up to 60 keV.

\begin{table}
	\centering
	\caption{4U2206+54 observations. }
        \begin{threeparttable}
	\label{tab:cyclobs}
	\begin{tabular}{lcccc} 
		\hline
		Instrument & Year & $F_{2-10}$\tnote{a}   & C.L. & Ref.  \\
		\hline
		\rxte & 1997 & 2.5 & N & 1 \\
                \sax & 1998 & 0.4 & Y & 1\\
                \rxte & 2001 & 1 & Y & 1 \\
                \integral (Rev. 67)& 2003 & 15.9\tnote{b} & Y & 2\\
                \integral (Rev. 87) & 2003 & 6.3 & Y & 2\\
                \rxte & 2007 & 2.5 & N & 3\\
                {\it XMM} + \integral & 2011 & 3.5 & N & 4\\
                \nustar & 2016 & 4 & N & 5 \\
		\hline
	\end{tabular}
              \begin{tablenotes}
              \item[a] Unabsorbed $2-10$ keV in units of $10^{-10}$ erg s$^{-1}$ cm$^{-2}$.
              \item[b] 4- 150 keV
               \item[1] \cite{2004A&A...423..301T}
               \item[2] \cite{2005A&A...438..963B}
               \item[3] \cite{2009A&A...494.1073R}
               \item[4] \cite{2012MNRAS.425..595R}
                \item[5] This work
              \end{tablenotes}
            \end{threeparttable}
\end{table}



\section{Discussion}
\label{sec:disc}

The spin period of \so is rapidly decaying at a very high rate of
$\dot \nu=-1.8\times 10^{-14}$~Hz s$^{-1}$. This suggests
  a strong negative torque applied to the magnetosphere of a slowly
  rotating neutron star. Under the assumption of a normal
  field strength ($B\sim 10^{12}$ G) the magnetic
accretion model \citep{2013ARep...57..287I} is able to explain the
observed $\dot \nu$ provided that a) the stellar wind of the donor
star presents some level of magnetization ($B>70$ G) and b) a dense magnetic slab of ambient matter forms around the
magnetosphere of the NS. This disk like structure requires very
special conditions to form in a wind fed system and the model
parameters have to be within certain narrow ranges 
\citep[see][for details]{2013ARep...57..287I}. 

Alternatively, such
  a torque can arise at the stage of subsonic settling accretion
  \citep{2012MNRAS.420..216S} in a
  wind-fed X-ray binary, when a hot convective shell forms above the
  slowly rotating magnetosphere. The settling
  accretion requires an X-ray pulsar luminosity, $L_{\rm x}$, below a critical value of $\sim 4\times 10^{36}$ erg
  s$^{-1}$, which is the case for 4U 2206+54. The torque is caused by
  turbulent viscosity in the shell and can be negative if the X-ray
  luminosity drops down below the equilibrium value, $L_{\rm eq}$,
  determined by the balance between the angular momentum supply to, and
  removal from, the magnetosphere \citep[see][for
  more detail and relevant formulas]{2012MNRAS.420..216S}. The equilibrium luminosity
  depends on the NS magnetic dipole moment $\mu$ (the NS surface
  magnetic field $B=2\mu/R^3$, where $R$ is the NS radius), the binary
  orbital period $P_{\rm orb}$, the pulsar spin period $P_{*}$ and on
  the stellar wind velocity $v_{\rm w}$. In the case of 4U 2206+54, two
  possible binary orbital periods have been proposed 9.5 d
  \citep{2006A&A...449..687R} and 19 d \citep{2007ApJ...655..458C}. On
    the other hand, the magnetic field of the NS is unknown but the observed strong spin-down suggests the pulsar is not in equilibrium. Therefore, it is incorrect to use the equilibrium formulas to evaluate the NS magnetic moment. However, at X-ray luminosities much smaller than the equilibrium value, at the spin-down stage, it is still possible to obtain the lower limit of the NS dipole magnetic moment by neglecting the spin-up torque, which is independent of the stellar wind velocity and orbital binary period \citep[see formulas in][]{2017arXiv170203393S}:

\begin{equation}
\label{e:mu_lim1}
\begin{matrix}
\mu\ge\mu^{\prime\prime}\approx 1.7\times 
10^{30}[\mathrm{G\,cm}^3]\Pi^{\prime\prime}\left|\frac{2\piup
    \dot\nu}{10^{-12}\mathrm{Hz\,s}^{-1}}\right|^{11/13}\\
\left(\frac{L_x}{10^{36}\mathrm{erg\,s}^{-1}}\right)^{-3/13}
\left(\frac{P*}{100\mathrm{\,s}}\right)^{11/13}
\end{matrix}
\end{equation}

where $\Pi^{\prime\prime}\gtrsim 1$ is a combination of the dimensionless
theory parameters. For 4U2206+54 we thus obtain
$\mu>10^{31}$~G~cm$^3$, corresponding to the NS surface magnetic field
$B_{\rm s}>2\times 10^{13}$~G. On the other hand, another estimate can be made if the orbital
binary period $P_{\rm orb}$ and stellar wind velocity $v_w$ are used:

\begin{equation}
\label{e:mu_lim2}
\begin{matrix}
\mu\ge\mu^{\prime}\approx 0.94\times
10^{30}[\mathrm{G\,cm}^3]\Pi^{\prime}\left|\frac{2\piup
    \dot\nu}{10^{-12}\mathrm{Hz\,s}^{-1}}\right|^{1/2}\\
\left(\frac{v_w}{1000\mathrm{km\,s}^{-1}}\right)^{-3/2}
\left(\frac{P*}{100\mathrm{\,s}}\right)^{7/8}
\left(\frac{P_{\rm orb}}{10\mathrm{\,d}}\right)^{-3/8}
\end{matrix}
\end{equation}

Here $\Pi^{\prime}\gtrsim 1$ is another combination of dimensionless
theory parameters. Note that this estimate does not depend on the NS
X-ray luminosity because it is derived under the assumption that the
observed $\dot \nu$ corresponds to the maximum possible spin-down rate
of a NS at the settling accretion stage. Assuming $P_{\rm orb}=9.5$ d \citep{2014AN....335.1060S},
and using $v_w=350$~km~s$^{-1}$ \citep{2006A&A...449..687R}, we obtain $\mu>2\times
10^{32}$~G~cm$^3$, which corresponds to a
surface magnetic field $B_{\rm s}>2\times 10^{14}$~G. From both calculations,
we can estimate a lower limit for the surface field strength $B_{\rm s}$ of several
times $10^{13}$~G. The lack of a CRSF detection with \nustar would be
consistent with this scenario. This
value is in the vicinity of the quantum critical field\footnote{$B_{\rm
  cr}=\frac{m_{e}^{2}c^{2}}{\hslash e}=4.4\times 10^{9}$ T, at which the
cyclotron energy of the electron, $\hslash \omega_{c}$ equals its rest
mass energy $m_{e}c^{2}$} $B_{\rm
  cr}=4.4\times 10^{13}$ G, which traditionally has been used to
delimit the high end of the magnetic field
distribution for pulsars from magnetars
\citep[][Fig. 7]{2014ApJS..212....6O}. Thus, among the wind accretion
powered X-ray pulsars, \so
would harbour a very high magnetic field NS.

\section{Conclusions}
\label{sec:conc}

From our analysis, the following conclusions can be drawn: 

\begin{itemize}
\item \nustar spectra rule out any cyclotron line up to 60 keV.
\item The secular strong spin down of \so is confirmed, at a rate of $\dot \nu=-1.8\times 10^{-14}$~Hz s$^{-1}$ 
\item Under the spherical settling accretion scenario, the required
  surface magnetic field needs to be, at least, several times $10^{13}$ G,
  at the high end of the magnetic field distribution for pulsars.  Thus, \so
appears as a strongly magnetized NS, whose X-ray emission is powered by 
the accretion of the slow wind of its main sequence (O9.5V).  
\end{itemize}

\section*{Acknowledgements}

This research has been supported by the grant ESP2017-85691-P. This research has made use 
of a collection of \textsc{isis} functions (\textsc{isisscripts}) provided by 
ECAP/Remeis observatory and MIT 
(\texttt{http://www.sternwarte.uni-erlangen.de/isis/}). PR thanks
K. Kovlakas for fruitful discussions about the use of the Lomb-Scargle
periodogram and its implementation in \texttt{Python}. KP acknowledges
support from the RFBR grant 18-502-12025. The authors acknowledge the
anonymous referee whose constructive criticisim greatly improved the
presentation of the paper. 




\bibliographystyle{mnras}
\bibliography{bibliografia} 







\bsp	
\label{lastpage}
\end{document}